\documentclass[11pt]{article}

\usepackage{amsmath}
\usepackage{algpseudocode}
\usepackage{multirow}
\usepackage{enumitem,relsize}
\usepackage{graphicx,placeins}
\usepackage{algorithmicx}
\usepackage{algorithm}
\usepackage{algpseudocode}
\usepackage[font={sf}]{caption}
\usepackage{setspace,url}
\usepackage{microtype,cite,xfrac}
\usepackage[letterpaper]{geometry}
\usepackage{amsthm}
\usepackage{framed}\usepackage[svgnames]{xcolor}
 
 \colorlet{shadecolor}{blue}   
 \renewenvironment{leftbar}{%
   \MakeFramed {\advance\hsize-\width \FrameRestore}}%
{\endMakeFramed}

\newenvironment{satoshi}
   {\begin{leftbar}\begin{quote}\color{blue}}
   {\end{quote}\end{leftbar}}
   
\usepackage{xspace}

\newcommand{\para }[1]{\smallskip \noindent {\bf #1}}

\newcommand{\Q}{{ Q}}

\algnewcommand\algorithmicinput{\textbf{INPUT:}}
\algnewcommand\INPUT{\item[\algorithmicinput]}
\newcommand\numberthis{\addtocounter{equation}{1}\tag{\theequation}}

\title{\textbf{An Explanation of Nakamoto's\\ Analysis of Double-spend Attacks}}

\author{A.~Pinar Ozisik 
and Brian Neil Levine\\
College of Information and Computer Sciences\\
University of Massachusetts Amherst} 

\date{}

\makeatletter
\def\blfootnote{\xdef\@thefnmark{}\@footnotetext}
\makeatother

\begin{document}

\maketitle
 
\begin{abstract}
The fundamental attack against blockchain systems is the double-spend
attack. In this tutorial, we provide a very detailed explanation of
just one section of Satoshi Nakamoto's original paper where the
attack's probability of success is stated. We show the derivation of
the mathematics relied upon by Nakamoto to create a model of the
attack. We also validate the model with a Monte Carlo simulation, and
we determine which model component is not perfect.

\end{abstract}

\section{Introduction}
\blfootnote{This work's copyright  is owned by the authors. A non-exclusive license to distribute the pdf has been given to arxiv.org.}
Satoshi Nakamoto's whitepaper\cite{Nakamoto:2009} introduced a protocol for distributed consensus using blockchains and also applied to a digital currency called Bitcoin. Nakamoto identified the primary attack against blockchain consensus: the double spend attack. The paper also analyses the chances of the attack's success, spending a couple pages to present the results.  

In this tutorial, we provide a very detailed explanation of that one section of Nakamoto's paper. We assume the reader has already read Nakamoto's paper at least once, and has a basic understanding of the blockchain algorithm for distributed consensus.  We show the derivation of the math relied upon by Nakamoto (and point out a small error). We also validate the equations with a  {\em Monte Carlo} simulation, and we determine which component of the model is not perfect.

We assume the reader is familiar with some mathematics related to probability. Because our intentions are to provide tutorial, our derivations are complete, eliding only the smallest algebraic steps. 

\para{The double spend attack.} To start, we review how the attack works. Say that we are a merchant that accepts Bitcoin in exchange for goods, such as a
cup of coffee or a car, or perhaps even dollars or some other virtual currency. We are worried about a malicious customer who
intends to use this vulnerability in Nakamoto's algorithm to defraud us. 
To be successful, the attacker needs to be in control of some reasonable amount of mining power. We'll say that she has a fraction of all mining power $0<q<1$, with the set of honest miners in control of the remaining fraction $p=1-q$.

We'll assume that the latest block on the blockchain is block $B_0$.
The attacker follows these steps:
\begin{enumerate}
\item The attacker sends a transaction $\cal{H}$ to the Bitcoin network
  that moves coin from an address that she controls to an address that the merchant controls. 
\item The merchant waits for $\cal{H}$ to appear on the
  blockchain in a block $B_1$, which has block $B_0$ as its previous, and possibly a follow-on sequence of blocks $B_2, \ldots, B_z$ to appear, and only then  moves to the next step. In other words, the merchant waits for a block $z$,  where $z \geq 1$.
\item The merchant hands over the  goods to the attacker.
\item The attacker releases a chain of blocks, $B'_1, \ldots, B'_{z+1}$. Block $B'_1$ has block $B_0$ as its previous. Contained in $B'_1$ is transaction $\cal{F}$, which  moves all the coin from the attacker's address to a second address in her control. 
\item If honest miners haven't reached block $B_{z+1}$ yet, the attacker wins. At that point, the miners will accept block $B'_1$ and the blocks the attacker mined afterwards; which means that the attacker has both the goods and her coin. 
\item If that attack hasn't won, she can continue attacking, using her mining power to race ahead 1 block more than the honest miners. 
\end{enumerate}

  The main question answered in this tutorial is: \textit{Given an attacker that controls a fraction $q$ of the mining power, and a merchant that
  waits for $\cal{H}$ to be $z$ blocks deep before
  releasing goods, what is the probability that the
  attacker can mine
  enough blocks to overtake the blockchain?}

\section{Probability of Double spend attack}

Let's explain  Satoshi's first thoughts on the attack~\cite{Nakamoto:2009}:
\begin{satoshi} 
  \color{blue} The race between the honest chain and an attacker chain
  can be characterized as a Binomial Random Walk. The success event is
  the honest chain being extended by one block, increasing its lead by
  +1, and the failure event is the attacker's chain being extended by
  one block, reducing the gap by -1.
\end{satoshi}
A {\em random walk} is a mathematical process that takes place along a series of states connected in a line. Each state is numbered, and we start from state 0. Flipping a coin, we move forward on heads, and backwards on tails, along a series of states.

Bitcoin is configured so that blocks are discovered about every ten minutes. In the double-spend attack, the attacker will generate a block on average every $10/q$ minutes, and the honest miners a block on average every $10/p$ minutes. But those are just averages. Because of the randomness inherent to mining, at any given moment, it's possible for the attacker to have generated a few more or a few less blocks than the honest miners. Our random walk will track the {\em difference} between their tallies.  

\bigskip
\centerline{\includegraphics[width=.7\columnwidth]{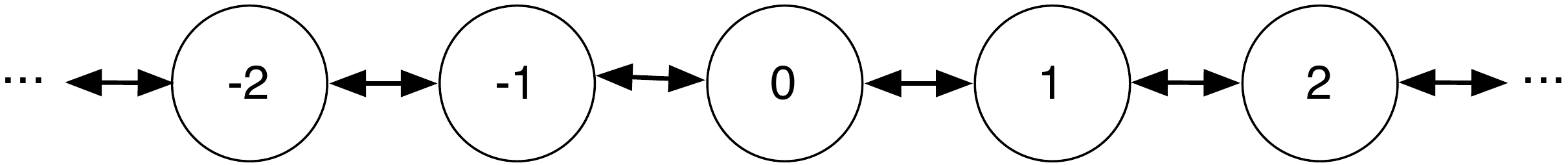}}\bigskip

\begin{satoshi}  \color{blue}
  The probability of an attacker catching up from a given deficit is
  analogous to a slight variation of the Gambler's Ruin
  problem. Suppose a gambler with unlimited credit starts at a deficit
  and plays potentially an infinite number of trials to try to reach
  breakeven. We can calculate the probability he ever reaches
  breakeven, or that an attacker ever catches up with the honest
  chain, as follows \cite{Feller:1968}:\\ \\
  $p$ = probability an honest node finds the next block \\
  $q$ = probability the attacker finds the next block \\
  $\Q_z$ = probability the attacker will ever catch up from $z$ blocks
  behind

 \begin{align}
  \Q_z=\begin{cases} 
    1  & \text{, if } p\leq q \\ 
    (\frac{q}{p})^z&  \text{, if } p>q \label{eq:satoshiqpz}\\
  \end{cases}
 \end{align}
\end{satoshi}
To understand the derivation of Eq.~\ref{eq:satoshiqpz}, for which Nakamoto cites a well-known 1968 textbook from Feller\cite{Feller:1968}, we must delve into the  Gambler's Ruin problem and its ``slight variation". 
By the way, we've changed the notation slightly here from Nakamoto to make things easier and consistent. What Nakamoto (and Feller) denoted as ``$q_z$'' in Eq.\ref{eq:satoshiqpz} in the original paper, we write as $\Q_z$ here.

\subsection{The Gambler's Ruin Problem}

This famous problem was first studied by Blaise Pascal and Pierre de Fermat in 1656~\cite{Edwards:1983}. It models a gambler who enters a casino to play a simple game of chance. She starts with initial fortune of $i$ dollars, and makes a series of bets. Each bet causes her to either win \$1 with probability $q$ or
lose \$1 with probability $p=1-q$. Winning or losing each is independent of all other bets. The
gambler's goal is to win
$N$ dollars before going bankrupt at \$0. If the gambler is bankrupt, he
can't gamble any longer because he doesn't have money to pay \$1 in
the case of a loss. Reaching either $N$ or $0$ ends the game.

Let $q_i$ denote the probability that the gambler wins after some
number of gambles when he initially starts with
$i$ dollars. We know that $q_0=0$ since that state ends the game as a loss.  To determine the value of $q_i$, we note that there are only two ways to win the game starting with $i$ dollars. Either:
\begin{itemize}
\item the gambler wins his bet with probability $q$ and then wins
again with $i+1$ dollars he has obtained  with probability $q_{i+1}$;  or
\item the gambler loses
his bet with probability $p$ and then wins with the remaining 
$i-1$ dollars with probability $q_{i-1}$. 
\end{itemize}
That is, we can define this problem as a {\em recurrence relation}, where the future depends on only the current state:
\begin{align}
q_i = (q)q_{i+1} + (p)q_{i-1}  \label{eq1}
\end{align}
Since $p+q=1$, we also know that  
\begin{align}
q_i = (p)q_i+(q)q_i,
\end{align}
which we can substitute into the left side of Eq.~\ref{eq1} to obtain
\begin{align}
 q_{i+1} - q_{i}  &= \frac{p}{q}(q_i - q_{i-1}).\label{eq:qi+1}
\end{align}
From Eq.\ref{eq:qi+1}, we can build up to a general case with specific instances of $i$:  
\begin{align}
q_2 - q_1 = \frac{p}{q}(q_1-q_0).
\end{align}
Next, we increment and substitute:
\begin{align}
q_3 - q_2 &= \frac{p}{q}(q_2-q_1) 
 		  = \Big(\frac{p}{q}\Big)^2q_1.
\end{align}
We can generalize this
observation as
\begin{equation}
  q_{i+1} - q_i = \Big(\frac{p}{q}\Big)^iq_1.
\label{eq2}
\end{equation}
We use Eq.~\ref{eq2} in this little mathematical trick below to obtain the following:
\begin{align}
q_{i+1} - q_1
  &= q_{i+1}+ (- q_i + q_{i}) + (- q_{i-1}+q_{i-1}) + ... +(-q_{3}+q_{3})+( -q_{2}+q_{2}) - q_1 \\
  &= (q_{i+1}-q_i) + (q_{i}-q_{i-1}) + ... + (q_{3}-q_{2}) + (q_{2}-q_{1})\\
  &= \sum^{i}_{k=1}(q_{k+1}-q_k) \\
  &= q_1\sum^{i}_{k=1}\Big(\frac{p}{q}\Big)^k \\
\intertext{In other words:}
q_{i+1} &=   q_1\sum^{i}_{k=0}\Big(\frac{p}{q}\Big)^k\label{eq5}
\end{align}
Eq.~\ref{eq5} is the sum of a geometric series, which can
 be written as \mbox{$\sum^{i}_{n=0}a^i = \frac{1-a^{i+1}}{1-a}$} for any
number $a$ and integer $i \geq 1$. Therefore, Eq.~\ref{eq5} can be
written as:
\begin{equation} 
 q_{i+1}=
  \begin{cases}
    q_1 \frac{1-(\frac{p}{q})^{i+1}}{1-(\frac{p}{q})} & \text{, if}\ p\neq q \\
    q_1(i+1) & \text{, if}\ p=q=0.5 \\
  \end{cases}\label{eq4}
\end{equation}
We are getting closer, but we still have our result in terms of $q_1$, so let's see if we can get rid of that term.  We know that $q_N = 1$ because the gambler has reached his goal. Rewriting Eq.~\ref{eq4} and setting each case equal to  1, we have 
\begin{equation}
 q_N=
  \begin{cases}
    q_1 \frac{1-(\frac{p}{q})^{N}}{1-(\frac{p}{q})} = 1 & \text{, if}\ p\neq q \\
    q_1(N) = 1 & \text{, if}\ p=q=0.5 \\
  \end{cases}
\end{equation}
We solve for $q_1$ in each case:
\begin{equation} \label{eq6}
 q_1=
  \begin{cases}
    \frac{1-(\frac{p}{q})}{1-(\frac{p}{q})^N} & \text{, if}\ p\neq q \\
    \frac{1}{N} & \text{, if}\ p=q=0.5 \\
  \end{cases}
\end{equation}
Eq.~\ref{eq6} can be plugged back into Eq.~\ref{eq4} to obtain:
\begin{align*} 
 q_{i+1}&=
    \begin{cases} \numberthis \label{eq8}
      \frac{1-(\frac{p}{q})^{i+1}}{1-(\frac{p}{q})^N} & \text{, if}\ p\neq q \\
      \frac{(i+1)}{N} & \text{, if}\ p=q=0.5 \\
  \end{cases} 
\end{align*}
That's it.  For any integer $i$, we can compute $q_i$:
\begin{equation} \label{eq7}
 q_i=
  \begin{cases} 
      \frac{1-(\frac{p}{q})^i}{1-(\frac{p}{q})^N} & \text{, if}\ p\neq q \\
      \frac{i}{N} & \text{, if}\ p=q=0.5 \\
  \end{cases}
\end{equation}
The equations for this section were adapted from notes by Prof.~Karl Sigman~\cite{Sigman:2016}.

\subsubsection{A Slight Variation of Gambler's Ruin}
To work our way back to Nakamoto, we need to alter the game. Nakamoto states the probability that the attacker ``will \emph{ever} catch up", which is a pretty determined attacker.  Nakamoto isn't analyzing whether the economics work out: it may be that the attacker spends more on mining than is recovered from the double spent transaction; or it may be that the {\em coinbase} reward from announcing a tremendous number of new blocks is worth more than the double spent transaction.  Nakamoto's goal is instead solely to analyze the worst-case scenario where the attacker spares no expense in running their existing mining power to win the Gambler's Ruin. 

To reach Nakamoto's goal, we'll first let the attack lose up to $y$ dollars before quitting (and then we'll see what happens when $y$ goes to infinity). 
 Therefore, this
slight variation is converted to the original Gambler's Ruin in the
following way: The gambler starts with
$i=y$ dollars and the game ends either at \$0, which is a loss, or at $N=y+z$ dollars,
which is a win. In that case, our assumptions still hold for Eq.~\ref{eq7}, that $q_0=0$ and $q_N=1$. By substituting into Eq.~\ref{eq7}, we get:
\begin{equation}
 q_l=
  \begin{cases}
      \frac{1-(\frac{p}{q})^y}{1-(\frac{p}{q})^{y+z}} & \text{, if}\ p\neq q \\
      \frac{y}{(y+z)} & \text{, if}\ p=q=0.5
  \end{cases}\label{eq:limited}
\end{equation}
Consider the case where gambler is willing to lose an infinite amount of money, and
therefore, has unlimited resources. In other words, where $y$ goes to infinity. In the case where $p<q$,
$(p/q)^y \to 0$ as $y \to \infty$:
\begin{align}
\lim_{y\to\infty}  \frac{1-(\frac{p}{q})^y}{1-(\frac{p}{q})^{y+z}} = 1 \text{, when } p<q
\end{align}
In the case where $p>q$, to calculate the limit, we pull $(\sfrac{p}{q})^y$ as factor from numerator and denominator:
\begin{align}
\frac{1-(\frac{p}{q})^y}{1-(\frac{p}{q})^{z+y}}  &= 
\frac{(\frac{p}{q})^y \Big((\frac{p}{q})^{-y} -1\Big)}{(\frac{p}{q})^y \Big((\frac{p}{q})^{-y} -(\frac{p}{q})^z\Big)}
=\frac{  (\frac{p}{q})^{-y} -1}{ (\frac{p}{q})^{-y} -(\frac{p}{q})^z}
\end{align}
When $p>q$, $(\frac{p}{q})^{-y} = (\frac{q}{p})^y \to 0$ as $y \to \infty$:
\begin{align}
\lim_{y\to\infty}  \frac{(\frac{p}{q})^{-y} -1}{ (\frac{p}{q})^{-y} -(\frac{p}{q})^z} = \frac{-1}{-(\frac{p}{q})^z} = \bigg(\frac{q}{p}\bigg)^z \text{, when } p>q\label{qzlim}
\end{align}
Because Eq.~\ref{qzlim} assumes that the attacker has unlimited resources, we can't use our existing notation (``$q_\infty$" doesn't really make sense), and so we'll switch notation and let $\Q_z$ denote the probability of catching up from a deficit of $z$ given unlimited resources:
\begin{equation} \Q_{\text{z}}=
  \begin{cases} 
      1 & \text{, if}\ p < q \\
      (\frac{q}{p})^z & \text{, if}\ p > q
  \end{cases}
 \label{eq:Qz} \end{equation}

\noindent The equations for this section were adapted
from notes by L.~Rey-Bellet~\cite{Rey-Bellet:2016}.

\subsection{Analogy to the Double-Spend Attack on a Blockchain}
The analogy to our blockchain scenario follows directly. Let $p$ be the mining power and probability that the honest miners find the
next block, and let $q$ be the attacker's mining power. We define $q + p = 1$, since we assume either only
the attacker or honest miners will win a block each round (and not some late comer or third-party).
If the attacker has unlimited resources for mining and stops when he
reaches $z$, then we can use Eq.~\ref{eq:Qz}.

\bigskip
It's worth noting an error by Nakamoto here. We aren't interested in $\Q_z$, the probability that the attacker will simply catchup. Instead, Nakamoto should have calculated $\Q_{z+1}$, the probability of the attacker going one past the honest miners.

\section{Poisson Experiments}

Satoshi continues with the following analysis.
\begin{satoshi} \color{blue}
  The recipient waits until the transaction has been added to a block
  and $z$ blocks have been linked after it. He doesn't know the exact
  amount of progress the attacker has made, but assuming the honest
  blocks took the average expected time per block, the attacker's
  potential progress will be a Poisson distribution with expected
  value:
\begin{align}
\lambda=z\frac{q}{p}
\end{align}
\end{satoshi} 

To find this expected value, Satoshi is using a mathematical model called a {\em Poisson
  experiment}.
In a Poisson experiment, we model a real situation involving probability by  counting the number of {\em successes} in a series of {\em intervals} measured in time. To use such a model, we must assume the following~\cite{walpole:2012}:
\begin{enumerate}
\item The number of successes during each time interval is independent of  any other   interval. 
\item The probability that a single success will occur during a very short time interval is proportional to the duration of the time interval.
\item The probability of more than one success in such a short time interval is negligible.
\end{enumerate}
(We are also assuming that the probability for success does not change during the experiment, though in reality, miners can increase or decrease their resources.) 

To make use of the well-known results for Poisson experiments, our
first job is to figure out a value $\lambda$, which is the average
number of successes we expect during each  interval. It's a
rate: successes/interval. For us, successes are the number of blocks
we expect the attacker to discover. And an interval is the time spent
by the merchant waiting for $z$ blocks to be discovered by the honest miners.  In other words, the units for successes/interval is blocks/interval.

The Bitcoin network is configured so that every $T=10$ minutes, 1 block is discovered with 100\% of the current mining power. 
For honest nodes,  every $T$ minutes, $p$ blocks are discovered. To produce $z$ blocks, they'll need an interval of 
\begin{align}
z \mbox{ blocks} \cdot \frac{T\mbox{ minutes}}{p \mbox{ blocks}} = \frac{zT}{p} \mbox{ minutes.}
\end{align}
For the attacker, $q$ blocks are discovered every $T$ minutes.
Therefore, during that interval, the attacker will produce blocks at a rate of 
\begin{align}
\lambda =\left(\frac{zT}{p}\mbox{ minutes/interval}\right)\cdot \frac{q\mbox{ blocks}}{T \mbox{ minute}} = \frac{zq}{p} \mbox{ blocks/interval.}
\end{align}
 $\lambda=zq/p$ is just an average. And in a trial of our
Poisson experiment, we'll draw from the Poisson distribution (i.e., we'll roll a
Poisson-shaped die) to see how many success actually happened. Let's
say that $X$ successes happened in a particular trial. The probability
that $X=k$ successes occurred during the interval, where $k\geq 0$, is known to
be:
\begin{align}
  P(X=k;\lambda ) = \frac{\lambda^k e^{-\lambda}}{k!}\label{eq:fish}
\end{align}
The above equation is called the Poisson density function\cite{walpole:2012}. We'd need a few more pages to derive the formula, so let's assume it's true for now. 

\para{Example.} Say that an attacker has $q=1/4$ mining power , and the targeted merchant requires $z=3$ before releasing goods. Question: What is the probability
that the attacker produces $k=2$ blocks while waiting for the goods
{\em and} then overtakes the blockchain? 

First, given $q$, we know $p=1-q$, and also that
$\lambda=zq/p=z/3$. The probability of the attacker producing $k=2$
blocks during an interval where $z=3$ blocks are produced by honest
nodes is
\[P(X=2;\lambda=z/3)=\frac{(3/3)^2 e^{-3/3}}{2!}=\frac1{2e}\approx
  0.18\]
Second, using Eq.~\ref{eq:Qz}, we know  the probability that the attacker will
eventually catch up to the $z-k$ blocks left is  $(q/p)^{z-k}$. Therefore in this case,
$(1/3)^{3-2} = 1/3$. We need the first and second parts to both be true, so we multiply: the answer to
the question is $(1/2e)(1/3) = 1/(6e) \approx 6\%$.

\subsection{The General Case}
We want to know the answer to a more general question: Given that the
merchant will wait for $z$ blocks before handing over physical goods,
what is the probability that the attacker with mining power $q$ can produce more blocks than the honest miners at that point or in the future? 

Satoshi's answer is as follows. Let $X$ be a random variable representing the number of blocks that the attacker discovers during the time that honest miners discover $z$ blocks. We already defined $P(X; \lambda)$ to be the probability of the attacker producing $X$ blocks. We know the probability  of catching up from the remaining $z-k$ difference is $q_{z-k}$. Therefore, to find the total probability of catching up, we sum all possibilities of $X$:
\begin{align}
&=P(X\!\!=\!0;\lambda)\Q_z + P(X\!\!=\!1;\lambda)\Q_{z-1}+\ldots \\
&=\sum^\infty_{k=0}P(X=k;\lambda)\Q_{z-k}\\
&=\sum_{k=0}^\infty \frac{\lambda^k e^{-\lambda}}{k!}  \left(\Q_{z-k}\right)
\end{align}

In fact, when $k>z$,  the probability that the attacker will catch up is 1. And so, as Satoshi states it:
\begin{satoshi} \color{blue} To get the probability the attacker could
  still catch up now, we multiply the Poisson density for each amount
  of progress he could have made by the probability he could catch up
  from that point:
\begin{align}
 \sum_{k=0}^\infty \frac{\lambda^k e^{-\lambda}}{k!} \cdot \left\{\begin{array}{lr}
(q/p)^{z-k} &\text{, if $k \leq z$}\\
1 &\text{, if $k > z$}
\end{array}
\right\}
\end{align}   
\end{satoshi} 
 
Last, since the probability that something happens is equal to the 1
minus the probability that it doesn't, Satoshi rearranges for our
(almost) final result. Here we subtract from 1 the probability that the attacker mines $k$ blocks and does {\em not} catch up.
\begin{align}
=&\sum_{k=0}^\infty \frac{\lambda^k e^{-\lambda}}{k!} \cdot \left\{\begin{array}{lr}
(q/p)^{z-k} &\text{, if $k \leq z$}\\
1 &\text{, if $k > z$}
\end{array}
\right\}\\
=& 1-\mathlarger{\sum}_{k=0}^\infty \frac{\lambda^k e^{-\lambda}}{k!} \cdot \left\{\begin{array}{lr}
1-(q/p)^{z-k} &\text{, if $k \leq z$}\\
1-1 &\text{, if $k > z$}
\end{array}
\right\}\\
=&1-\mathlarger{\sum}_{k=0}^z \frac{\lambda^k e^{-\lambda}}{k!}\cdot \left(1-(\frac{q}{p})^{z-k}\right)+\sum_{k=z+1}^\infty \frac{\lambda^k e^{-\lambda}}{k!} \cdot (0)\\
=&1-\sum_{k=0}^z \frac{\lambda^k e^{-\lambda}}{k!}\cdot \left(1-(\frac{q}{p})^{z-k}\right)&
\end{align}   
Or as Nakamoto states more succinctly: 
\begin{satoshi} \color{blue} Rearranging to avoid summing the infinite
  tail of the distribution...
\begin{align}
  1- \sum_{k=0}^z \frac{\lambda^k e^{-\lambda}}{k!}\left(1-(\frac{q}{p})^{z-k}\right)
\end{align}
\end{satoshi}

\begin{figure}[t]
\centerline{\includegraphics[width=.6\columnwidth]{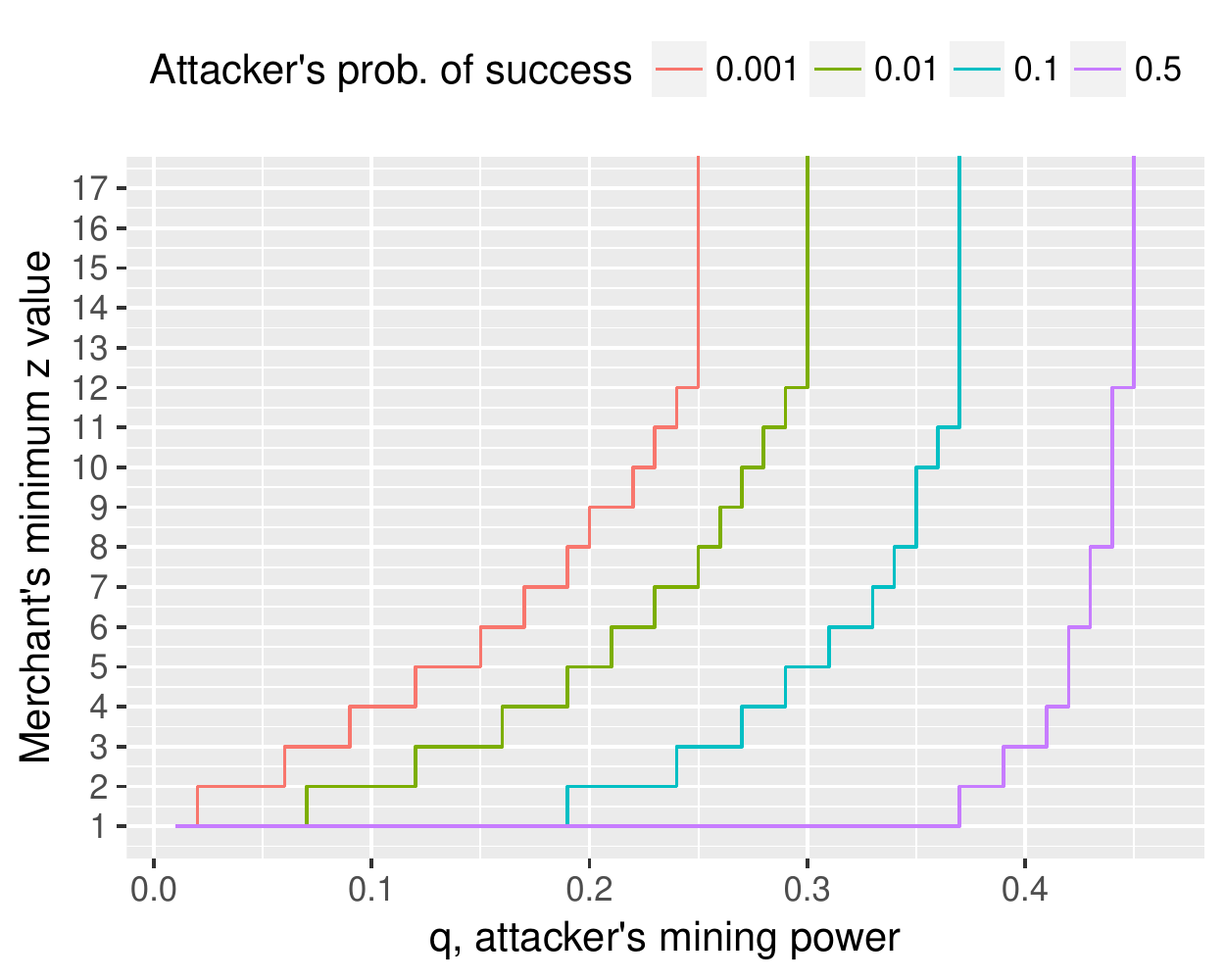}}
\caption{Given the attacker's mining power, and a desired probability of success (e.g, 0.001, 0.01, 0.1, and 0.5), the plot shows the minimal value of $z$ required of the merchant. Based on Eq.~\ref{eq:corrected-satoshi}.}\label{fig:q-vs-minz}
\end{figure}

\noindent Again, we must correct Nakamoto's error here. We are interested in the attacker surpassing the honest miners.
\begin{align}
=&  1- \mathlarger{\sum}_{k=0}^{z+1} \frac{\lambda^k e^{-\lambda}}{k!}\left(1-(\frac{q}{p})^{z+1-k}\right)\label{eq:corrected-satoshi}
\end{align}   

We aren't quite done. We want to know the minimum value $z$ such that
the probability of success is low, for example, just 1\%. Rather than
solving for that value, Satoshi provides us with a tiny bit of code
that enumerates all values and picks the minimum. Figure~\ref{fig:q-vs-minz} does the same for Eq.~\ref{eq:corrected-satoshi} as a plot rather than a list of values.

\para{Limitations.}
There are a few limitations we need to be aware of. First of all, it's a challenge for the merchant to determine the attacker's mining power $q$. It's always possible that a previously unknown miner or an existing one could dedicate new resources to double spend attacks. The Fischer, Lynch, and Paterson (FLP)~\cite{Fischer:1985} impossibility result tells us that Satoshi's algorithm can never reach consensus. At any point in time, the last block on the chain is only an estimate of what consensus will be in the future. 

Second, the model is a little strange in that the budget for the attacker is infinite for playing the Gambler's Ruin; yet, the attacker has limited computational power. Satoshi was perhaps attempting to quantify the worst case for each value of $q$; but in reality, the worst case is simply any value of $q>0.5$.
It's like saying you have infinite money for gas for your car, but can't spend any of those funds on a faster car, even though faster cars are available.

\section{Validation}
 
\begin{figure}[t]
\begin{minipage}[t]{.49\columnwidth}
\centerline{\includegraphics[width=\columnwidth]{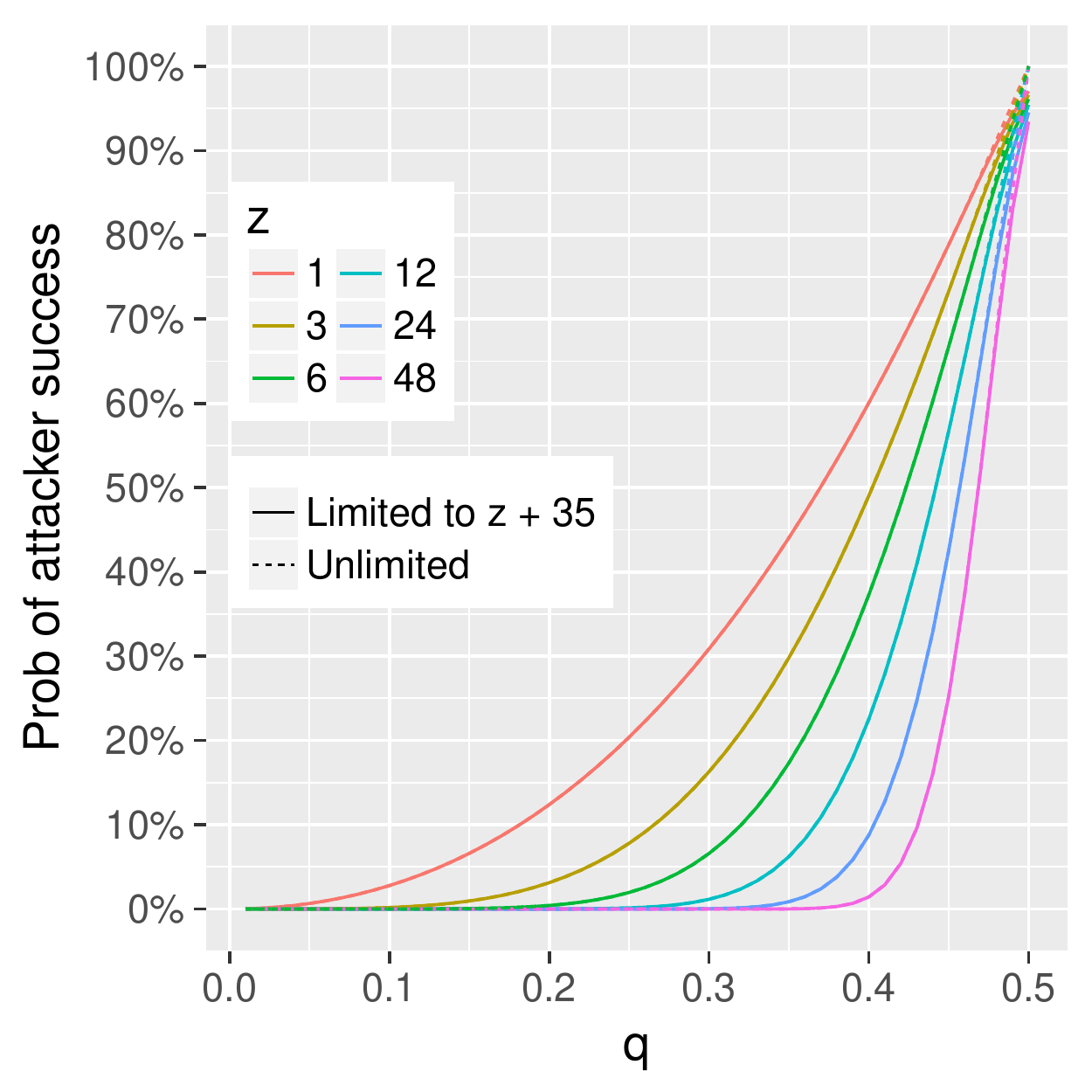}}
\caption{Eqs.~\ref{eq:corrected-satoshi} (an infinite budget for $\Q_{z+1}$) and Eq.~\ref{eq:z35} (a budget of $y={z+35}$) plotted for particular values. Here we see that a limited budget is almost the same result when $q<0.45$ or so.}\label{fig:z35}
\end{minipage}\hspace{.5cm}
\begin{minipage}[t]{.49\columnwidth}
\centerline{\includegraphics[width=\columnwidth]{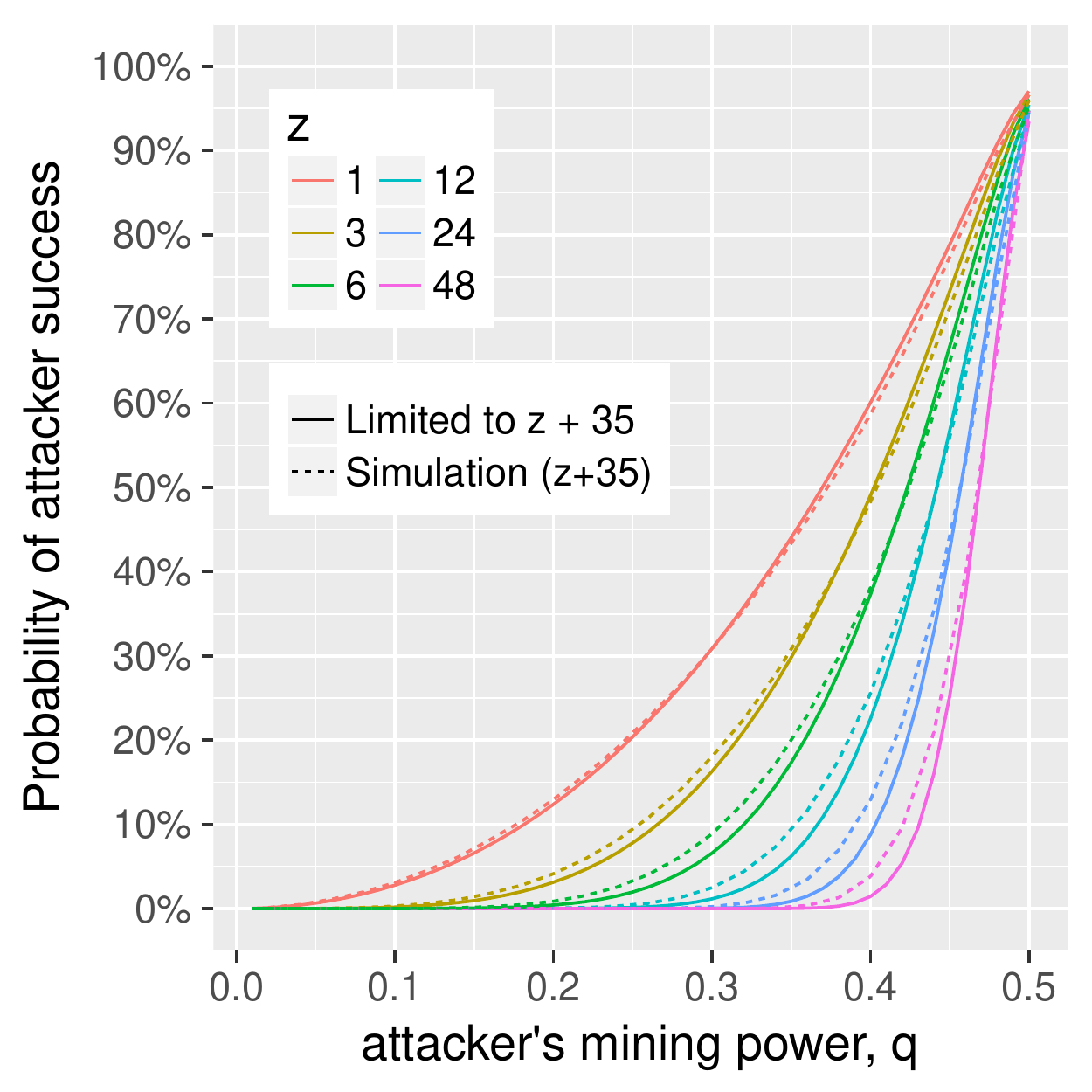}}
\caption{A comparison of the model (Eq.~\ref{eq:z35}) to the Monte Carlo simulation. The model and the simulation don't align exactly, but are a good match. Not bad!}\label{fig:val}
\end{minipage}
\end{figure}

Noticeably missing from Nakamoto's whitepaper is validation of the model.  To do that, we need to simulate the attack using a separate method of analysis. We'll use {\em Monte Carlo} simulation. We wrote a short Python script to simulate mining and the Gambler's Ruin race between honest miners and the attacker. The program does not use any of the equations from the model to predict the winner in any given trial. The program does not actually mine coin, it simply flips some coins to see whether each miner wins a block as simulated time passes. Hence, many trials can be run in a short period of time.

We can't simulate the case of an attacker with infinite resources. We need to find a finite case. Fortunately, a budget of $y=z+1+34$ is pretty close to the unlimited case, as Figure~\ref{fig:z35} illustrates. To model and plot the limited budget, we use this equation, based on Eq.~\ref{eq:limited}:
\begin{align}
=&  1- \mathlarger{\sum}_{k=0}^{z+1} \frac{\lambda^k e^{-\lambda}}{k!}\left(1- \frac{1-(\frac{p}q)^{z+35-k}}{1-(\frac{p}q)^{(z+35-k+z+1-k)}}\right)
\label{eq:z35}
\end{align} 
 
\begin{figure}[!htbp]
\includegraphics[width=.49\columnwidth]{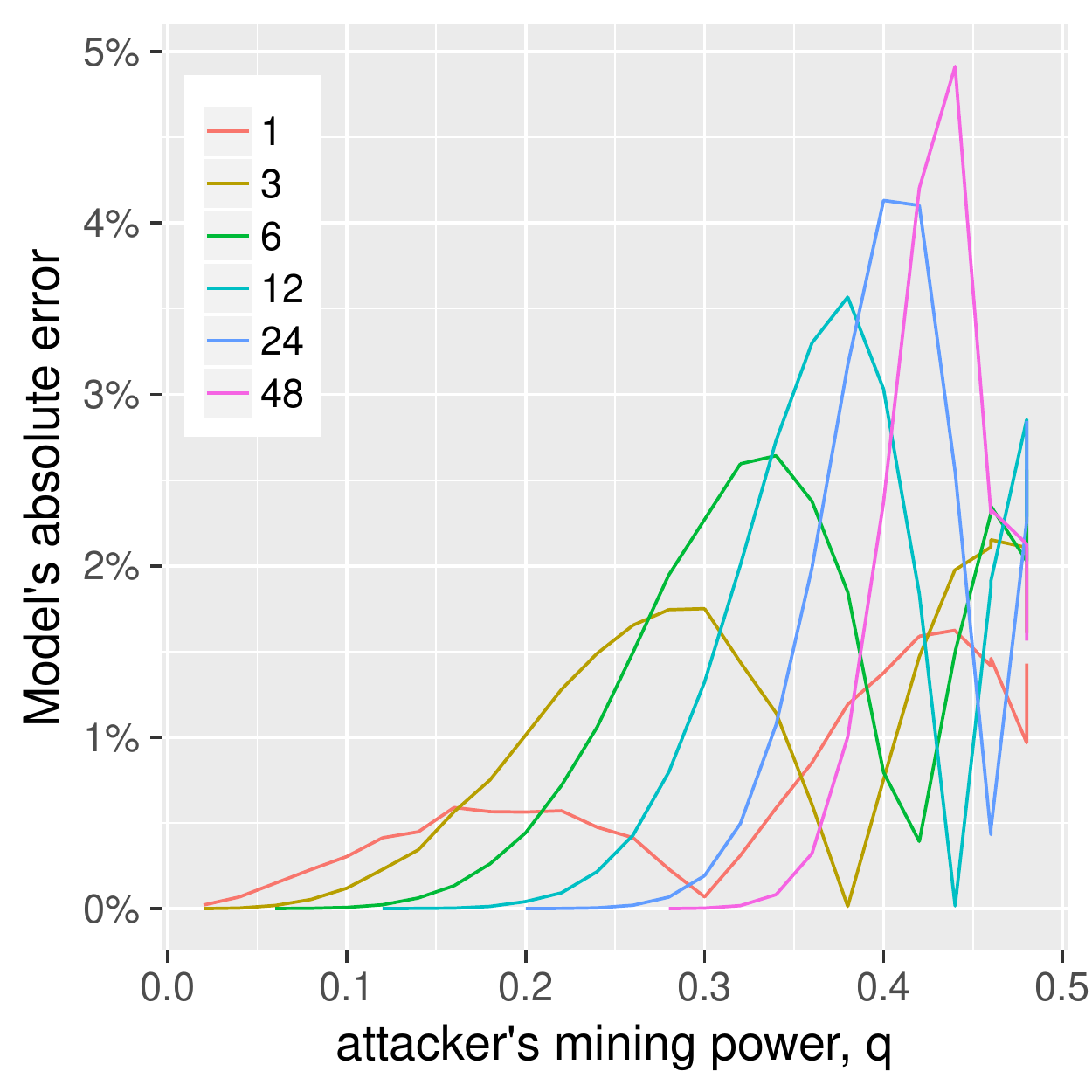}
\includegraphics[width=.49\columnwidth]{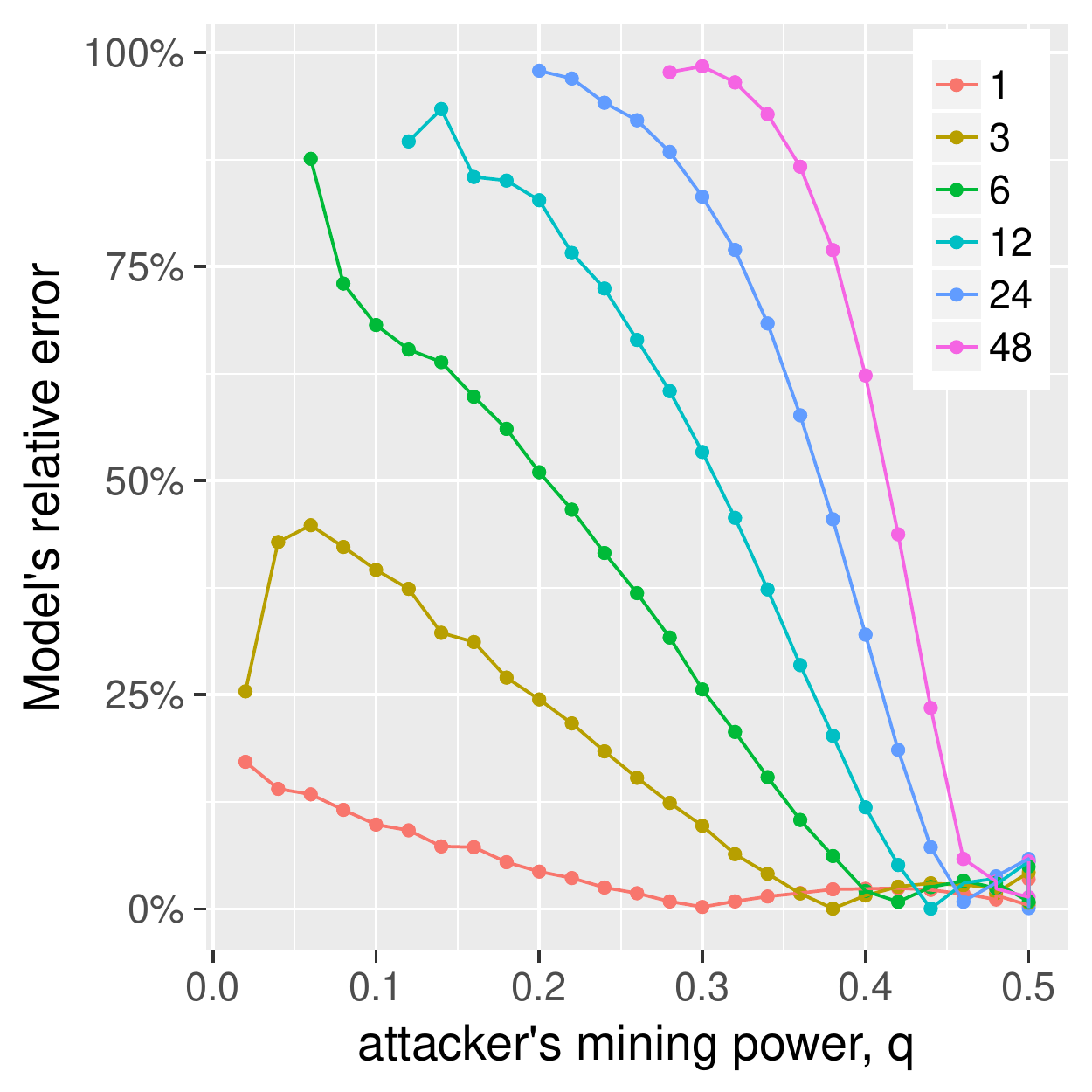}
 \caption{ The error of the model (Eq.~\ref{eq:z35}) against a simulation.  (Left)~Absolute error; (Right)~Relative error. The values for relative error are a little tremulous for low values of $q$ because winning events for the attacker are incredibly rare. Additional simulations would help smooth out the values. }\label{fig:val-error}
 \end{figure}

Figure~\ref{fig:val} compares the results from the Monte Carlo simulation (with budget $y=z+35$) to the values predicted by the model. The model isn't a perfect match to the Monte Carlo simulation.
Figure~\ref{fig:val-error}(left) shows the absolute error: under 4\% for all values of $z\leq 24$.   Figure~\ref{fig:val-error}(right) computes the relative error of the model with the simulation: as much as 100\% for lower values of $q$. Let's explore the source of the error.

\begin{figure}[!htbp]
\includegraphics[width=.49\columnwidth]{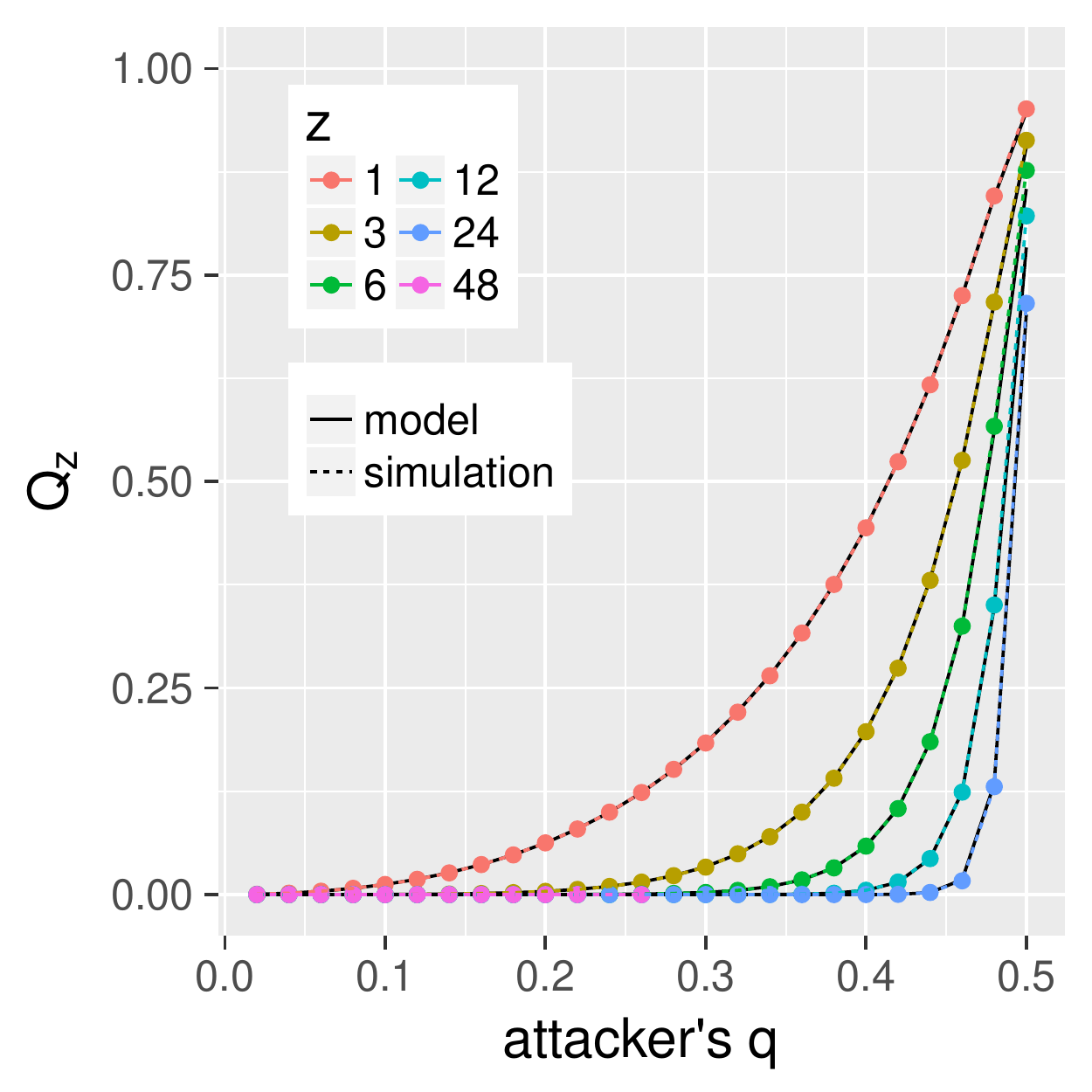}
\includegraphics[width=.49\columnwidth]{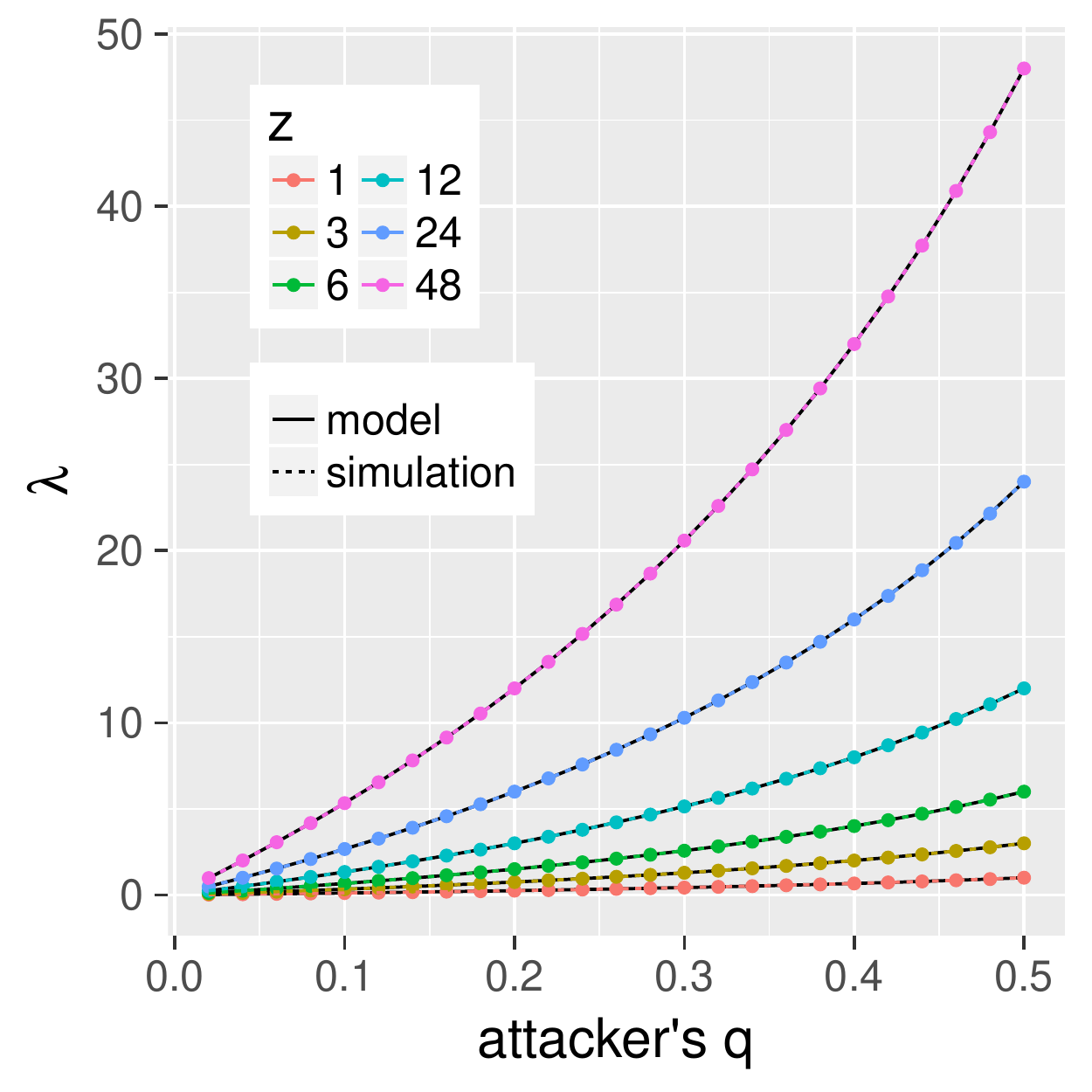}
\caption{(Left) The estimation of $\Q_{z+35}$ is not a source of error. (Right) The estimation of $\lambda$ is not the source of error.}\label{fig:valqz} \label{fig:vallambda}
\end{figure}

\begin{figure}[!htbp]
\centerline{\includegraphics[width=\columnwidth]{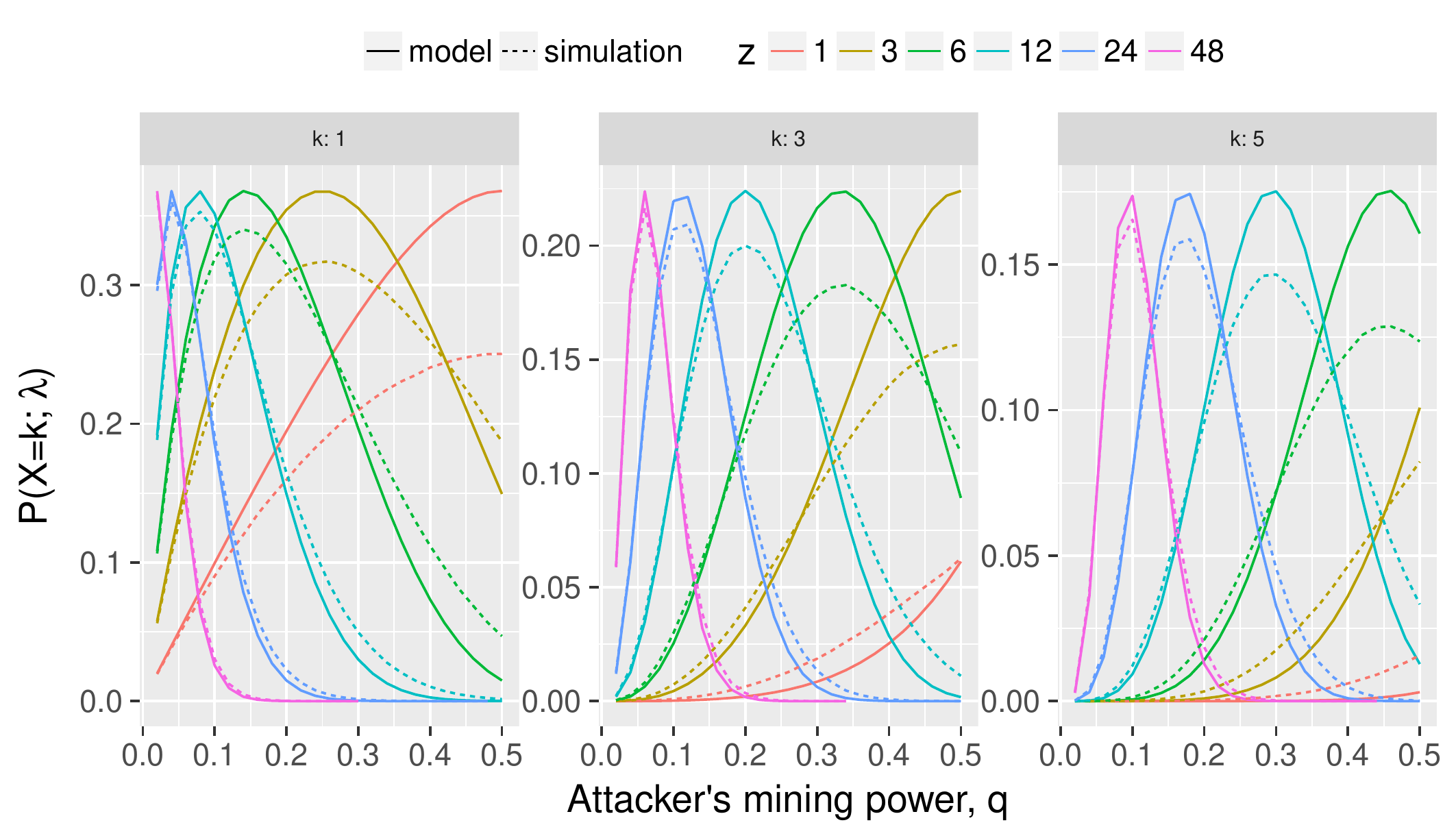}}
\caption{The Poisson density equation is the primary source of error. Example values from the Monte Carlo simulation and  values predicted by Eq~\ref{eq:fish} are shown. }\label{fig:valprobk}
\end{figure}

There are three components to the model: $Q_z$, $\lambda$, and the Poisson density function. 
Figure~\ref{fig:valqz}(Left) shows a comparison of the value of $Q_z$ from the simulation and the model, revealing a perfect match. Figure~\ref{fig:vallambda}(Right) compares the value of $\lambda$ from the simulation and the model, which are again a perfect match. Finally, we see in Figure~\ref{fig:valprobk} that the Poisson density function is  the source of error, especially for smaller values of $z$. 

\FloatBarrier
\bibliographystyle{abbrv}
\bibliography{references}

\end{document}